\documentclass[lettersize,journal]{IEEEtran}
\usepackage{amsmath,amsfonts}
\usepackage{algorithmic}
\usepackage{algorithm}
\usepackage{array}
\usepackage[caption=false,font=normalsize,labelfont=sf,textfont=sf]{subfig}
\usepackage{textcomp}
\usepackage{stfloats}
\usepackage{url}
\usepackage{verbatim}
\usepackage{graphicx}
\usepackage{cite}
\usepackage{color}
\usepackage{fancyhdr}

\begin{document}
\title{Toward Autonomous O-RAN: A Multi-Scale Agentic AI Framework for Real-Time Network Control and Management}

\author{Hojjat Navidan, Mohammad Cheraghinia, Jaron Fontaine,‌ Mohamed Seif, Eli De Poorter, H. Vincent Poor, Ingrid Moerman, Adnan Shahid
\thanks{Hojjat Navidan, Mohammad Cheraghinia, Jaron Fontaine, Eli De Poorter, Ingrid Moerman, and Adnan Shahid are with IDLab, Department of Information Technology at Ghent University - imec, Ghent, Belgium.}
\thanks{Mohamed Seif and H. Vincent Poor are with the Department of Electrical and Computer Engineering, Princeton University, USA.}}



\maketitle
\thispagestyle{fancy}

\begin{abstract}
Open Radio Access Networks (O-RAN) promise flexible 6G network access through disaggregated, software-driven components and open interfaces, but this programmability also increases operational complexity. Multiple control loops coexist across the service management layer and RAN Intelligent Controller (RIC), while independently developed control applications can interact in unintended ways. In parallel, recent advances in generative Artificial Intelligence (AI) are enabling a shift from isolated AI models toward agentic AI systems that can interpret goals, coordinate multiple models and control functions, and adapt their behavior over time. This article proposes a multi-scale agentic AI framework for O-RAN that organizes RAN intelligence as a coordinated hierarchy across the Non-Real-Time (Non-RT), Near-Real-Time (Near-RT), and Real-Time (RT) control loops.
(i) A Large Language Model (LLM) agent in the Non-RT RIC translates operator intent into policies and governs model lifecycles. (ii) Small Language Model (SLM) agents in the Near-RT RIC execute low-latency optimization and can activate, tune, or disable existing control applications; and (iii) Wireless Physical-layer Foundation Model (WPFM) agents near the distributed unit provide fast inference close to the air interface. We describe how these agents cooperate through standardized O-RAN interfaces and telemetry. Using a proof-of-concept implementation built on open-source models, software, and datasets, we demonstrate the proposed agentic approach in two representative scenarios: robust operation under non-stationary conditions and intent-driven slice resource control.
\end{abstract}

\begin{IEEEkeywords}

Open Radio Access Network, Ran Intelligent Controller, Agentic AI, Agentic RAN, 6G Networks, Intent-Driven Networking, 
\end{IEEEkeywords}

\section{Introduction}
\pagestyle{plain}

\IEEEPARstart{T}{he} transition toward 6G is being shaped by two converging revolutions: the disaggregation of the Radio Access Network (RAN) and the rise of generative Artificial Intelligence (AI). While Open-RAN (O-RAN) introduces flexibility through open interfaces and programmable control loops, the resulting network complexity exceeds the capabilities of traditional rule‑based optimization \cite{marinovaIntelligentORAN5G2024a}. Operators no longer manage monolithic network boxes; they orchestrate distributed, cloud-native RAN functions across multiple time domains, from the Service Management and Orchestration (SMO) layer to the RAN Intelligent Controller (RIC) and down to the Distributed Unit (O-DU). In emerging 6G settings where ultra-dense deployments, high mobility, multi-tenant slicing, and dynamic spectrum sharing become paramount, this complexity demands a new approach to RAN intelligence \cite{gkonisSurveyArchitecturalApproaches2025}.

To address this challenge, the community has increasingly adopted AI and Machine Learning (ML) solutions. Yet, current O-RAN intelligence is often realized as standalone rApps, xApps, and dApps that embed specialized, task-specific ML components \cite{hamdanRecentAdvancesMachine2023}. In practice, these components are commonly implemented as lightweight neural networks for physical layer inference, sequence models for traffic forecasting, or reinforcement learning agents for resource management and scheduling. While effective within their scope, these models are typically trained for a single objective and a fixed operating regime, and they are deployed as independent control or inference modules with limited awareness of network intent, peer actions, or longer-horizon effects. 

As operators introduce more standalone applications, unintended cross-model interactions emerge across the control hierarchy. For example, a handover-optimization xApp may increase the handover rate to improve mobility, but this can redirect users to cells that already have little spare capacity \cite{adamczykConflictMitigationFramework2023}. 
Without a coordinating layer that understands intent and manages cross-timescale dependencies, independently optimized xApps may conflict with one another, resulting in unstable control actions and degraded end-to-end performance \cite{zhangIntentDrivenClosedLoopControl2024}.


Several coordination mechanisms have been proposed within the O-RAN ecosystem, including conflict mitigation frameworks, multi-xApp arbitration layers, and digital twins to validate xApp behavior before deployment \cite{adamczykConflictMitigationFramework2023}. Although promising, these approaches are largely reactive, resolving conflicts only after different xApps issue competing control actions, and they typically arbitrate at the level of control variables without explicit awareness of higher-level intent. Consequently, they may prevent inconsistent parameter updates yet still fail to satisfy objectives such as slice isolation, latency guarantees, or operator policies. Moreover, they often assume models are static and manually curated, which limits their effectiveness in the face of evolving traffic patterns and model drift. 

At the same time, the AI ecosystem has begun to shift from models to agents, systems capable not only of prediction but also of planning, tool use, self-correction, and multi-step reasoning. Recent progress in Large Language Models (LLMs) has demonstrated their ability to interpret high-level goals, synthesize structured actions, and coordinate multiple specialized tools \cite{yaoReActSynergizingReasoning2022, schickToolformerLanguageModels2023}. In parallel, Small Language Models (SLMs) have emerged as efficient, edge-deployable alternatives capable of fast, domain-specific reasoning \cite{phamSlimLMEfficientSmall2025}. At the physical layer, Wireless Physical-Layer Foundation Models (WPFMs) utilize In-phase and Quadrature (I/Q) and Channel State Information (CSI) to pre-train general-purpose models. These models can be adapted to a range of downstream tasks, including interference detection, channel estimation, and localization, while meeting strict real-time constraints \cite{aboulfotouh6GWavesFMFoundation2025, cheraghiniaFoundationModelWireless2025}.

These developments present an opportunity to rethink O-RAN's intelligence stack. Instead of operating as isolated models, LLMs, SLMs, and WPFMs can serve as coordinated, intent-driven autonomous agents, each aligned with the appropriate RIC control loop. LLMs can reason over operator intents, network semantics, and long-term trends; SLMs can execute policy-driven control actions at Near-RT scales; and WPFMs can provide microsecond-scale inference at the PHY layer. What is missing is an architectural framework that elevates these heterogeneous AI components into a unified, multi-scale, agentic system capable of orchestrating decisions across Non-RT, Near-RT, and RT domains while maintaining safety, stability, and operator oversight.

This article bridges the semantic gap between high-level operational intent and radio-level control by proposing a hierarchical agentic AI framework for O-RAN. Specifically, we introduce an architecture that assigns LLM rApps to intent interpretation and governance in the Non-RT loop, SLM xApps to policy-driven optimization and xApp orchestration in the Near-RT loop, and WPFM dApps to real-time PHY inference close to the air interface. We further define the supporting data and telemetry pathways, model lifecycle mechanisms, and safety guardrails needed to make such cross-loop autonomy feasible in practice, and validate the approach through a proof-of-concept implementation for representative RAN control scenarios.


The remainder of this article is organized as follows. Section \ref{sec:framework} introduces the proposed agentic framework, outlining the high-level architectural goals, control loops, agent roles, data pathways, and safety mechanisms. Section \ref{sec:implementation} presents a proof of concept, describing the selected use cases, model choices, and evaluation methodology. Section \ref{sec:challenges} discusses the risks and challenges associated with deploying agentic intelligence in O-RAN. Finally, section \ref{sec:conclusion} concludes the article with final remarks and future directions.

\section{Agentic AI O-RAN Framework}
\label{sec:framework} 
In this section, we describe the system architecture and multi-timescale control loops, define the responsibilities of different components, and summarize the data telemetry pathways and safety mechanisms that support closed-loop governance across RIC layers.

\subsection{Architectural Overview and Control Loops}
\begin{figure}
    \centering
    \includegraphics[width=\linewidth]{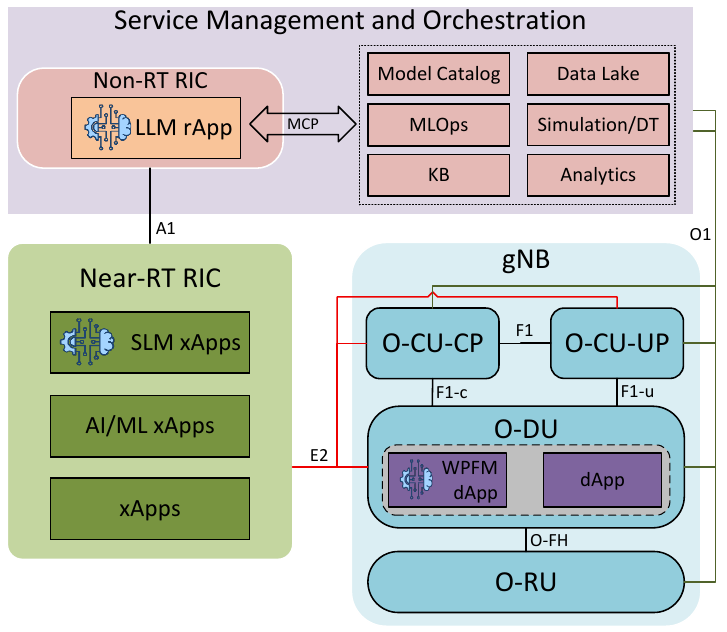}
    \caption{High-level architecture of the agentic AI framework.}
    \label{fig:arch}
\end{figure}

We propose the architecture in Fig. \ref{fig:arch} to enable collaborative, multi-scale intelligence within the O-RAN ecosystem. It builds on the disaggregated gNB split into the Centralized Unit (O-CU), O-DU, and Radio Unit (O-RU), interconnected through open interfaces such as F1 and O-FH. While this disaggregation enables flexible deployment and multi-vendor interoperability, it also highlights the inherently multi-timescale nature of RAN control. 
Accordingly, we map these tiers to a team of cooperating AI agents, with their cognitive capacity inversely proportional to their required response time: intent-level reasoning runs in the Non-RT loop, policy-driven control in the Near-RT loop, and fast neural inference in the RT loop, near time-critical functions.


The Non-RT RIC incorporates an LLM-driven rApp responsible for long-horizon reasoning, intent interpretation, policy generation, and model governance. This rApp interacts with a suite of SMO services via an internal Model Context Protocol (MCP) interface, enabling coordination between the language models and operational support systems. The Model Catalog maintains all registered SLM and WPFM variants together with metadata on their capabilities, constraints, and deployment history. The Data Lake aggregates multi-source telemetry, including O1 measurements, Key Performance Metric (KPM) logs, and offline training datasets, while the Knowledge Base (KB) stores domain rules, operator guidelines, and validated configuration patterns that help ground the LLM's reasoning.

The MLOps pipeline automates model versioning, validation, testing, and promotion, ensuring that new models can be safely onboarded or rolled back. Meanwhile, the Simulation/Digital Twin (DT) environment allows the LLM rApp to evaluate hypothetical configurations or test candidate SLM/WPFM behaviors before deploying them into the live network. Finally, the Analytics module provides long-horizon KPM processing, trend analysis, and performance scoring, allowing the rApp to assess the effectiveness of deployed policies and detect emerging model drift or degradation. Through these components, the Non-RT RIC gains the situational awareness and operational tooling needed to govern the full hierarchy of RAN intelligence.

At the Near-RT RIC, the architecture hosts a heterogeneous portfolio of control applications, including SLM-based xApps, conventional AI/ML model-based xApps, and legacy rule-based xApps. This layer translates high-level intent and policies received via the A1 interface into specific, context-aware radio control actions. In contrast to traditional xApps that are typically optimized for a single task, the SLM xApps introduced in this framework provide lightweight reasoning capabilities. These capabilities enable them to interpret semantic policy objectives, fuse multiple KPM streams, and coordinate their actions with other xApps while operating within the control loop's latency constraints. Control decisions generated at this layer, such as slice-level resource allocation, handover parameter tuning, or load balancing directives, are enforced on the RAN via the E2 interface, thereby closing the Near-RT control loop.

In parallel, the gNB domain follows the O-RAN functional split into O-CU-CP (Control Plane), O-CU-UP (User Plane), and O-DU, interconnected through the F1 interface family. The O-CU hosts higher-layer protocol processing and control-plane functions, while the O-DU executes time-critical MAC and PHY operations and interfaces with the O-RU over the O-FH fronthaul. Within the O-DU, we introduce WPFM-enabled dApps that perform ultra-low-latency inference directly on IQ samples or CSI features. These dApps provide fast PHY-layer intelligence, such as interference classification, channel estimation, beam management, or modulation guidance, while respecting the strict real-time constraints of the lower RAN. Conventional dApps or baseline schedulers remain colocated as a fail-safe fallback, ensuring robust operation in case of model degradation or timing violations.

Finally, to ensure the system supports self-monitoring and continuous learning, the O1 interface provides continuous feedback from managed elements to the SMO, streaming performance counters, fault logs, and trace data into the Data Lake. This telemetry provides the evidence required by the LLM rApp and the MLOps pipeline to evaluate deployed agents, detect performance regressions, and trigger model updates or rollbacks when needed. 


It is important to note that the real-time control loop is not yet fully standardized within the O-RAN Alliance. Nevertheless, emerging proposals and early implementations suggest placing real-time dApps directly inside or adjacent to the O-DU to meet microsecond-scale processing deadlines \cite{lacavaDAppsEnablingRealtime2025, doroDAppsDistributedApplications2022a}. Following this direction, our framework assumes that WPFM-driven dApps execute within the O-DU processing pipeline, keeping RT agents close to the air interface to deliver deterministic real-time intelligence.



\begin{figure*}[t]
    \centering
    \includegraphics[width=\linewidth]{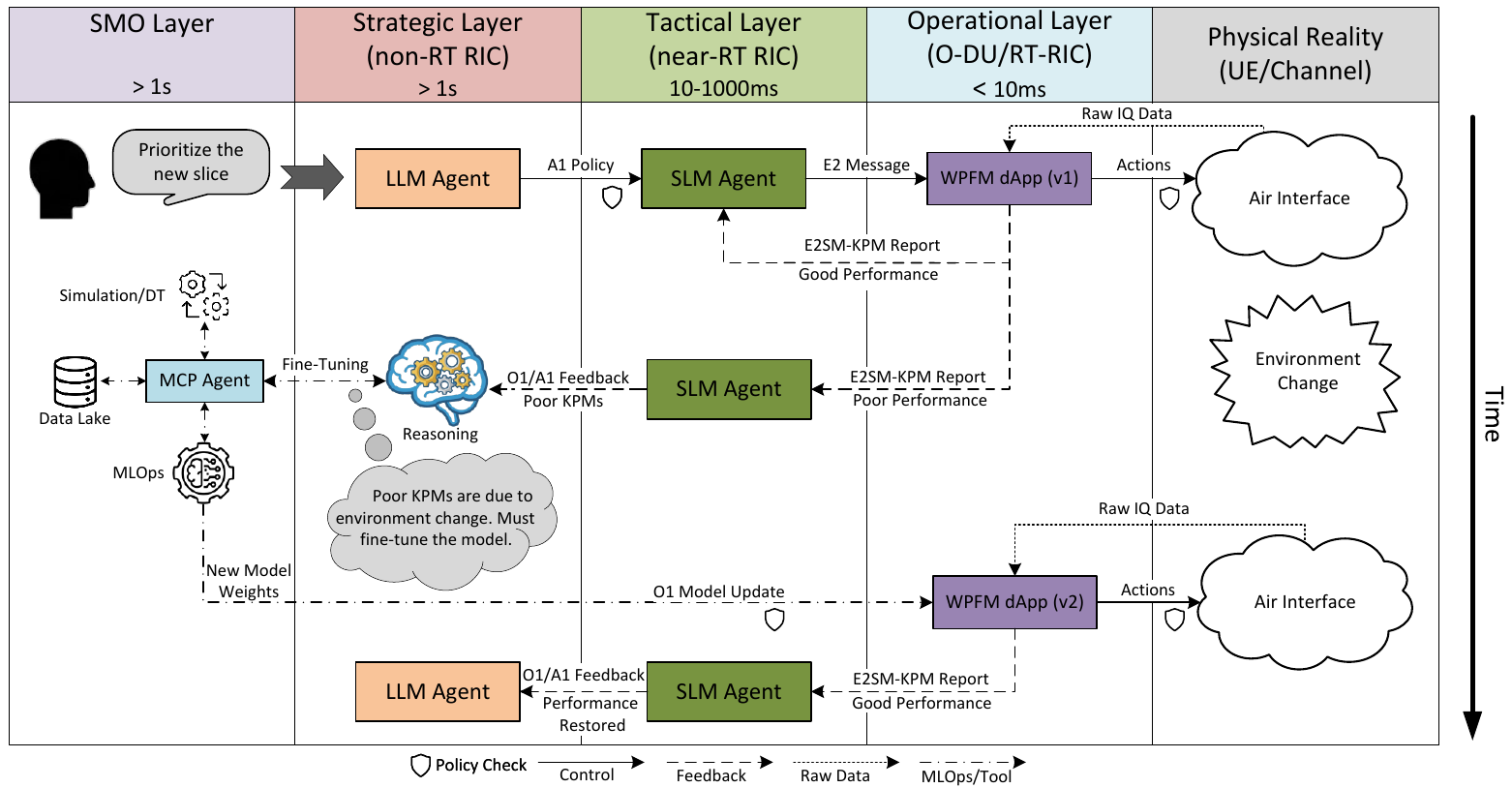}
    \caption{The interaction cycle in the proposed agentic O-RAN architecture. Operator intent is translated by the LLM agent into A1 policies, executed by SLM agents via E2 control, and realized through WPFM dApps that act on the air interface.}
    \label{fig:flow}
\end{figure*}

\subsection{Agent Roles, Data Flows, and Telemetry}
Our proposed system organizes agentic intelligence across three layers: strategic, tactical, and operational, using standardized O-RAN interfaces to connect them. The resulting interaction cycle, illustrated in Fig.~\ref{fig:flow}, demonstrates how intent, actions, feedback, and model evolution propagate across time and across the architecture.


The LLM rApp functions as a planner and long-horizon decision-maker in the strategic layer. It interprets operator intent, evaluates network-wide performance trends, and formulates structured A1 policies that encode high-level objectives such as prioritizing a slice during congestion or introducing new radio features. These policies are delivered to the Near-RT RIC, where they steer tactical behavior by defining resource floors and ceilings, priority relationships, and permissible update rates. The LLM rApp continuously monitors network behavior through aggregated telemetry and KPM summaries in the Data Lake, enabling intent-aware governance and performance tracking over time. Additional context is drawn from the Knowledge Base, while the Model Catalog maintains available model versions and their operational metadata. 

At the tactical layer, an SLM-based xApp serves as the primary execution agent for A1 intent. Beyond producing its own control decisions, the SLM can coordinate other applications, whether AI/ML-based or legacy rule-based, by selectively activating, deactivating, or reconfiguring them as policies and conditions change. Control actions are applied through E2 using the E2SM-RC service model to adjust parameters such as handover thresholds, resource allocations, and power-control settings. The outcomes of these actions are observed through E2SM-KPM reports, which provide Near-real-time performance feedback and are forwarded to the Data Lake for strategic analysis. This closed-loop interaction enables the Near-RT RIC to behave as a unified tactical layer rather than a collection of independent controllers. 

The operational layer resides within the O-DU and executes at microsecond-to-millisecond granularity. Here, dApps perform real-time inference over IQ samples and CSI, generating PHY-level insights that guide scheduling, beam selection, interference management, and link adaptation. These dApps operate under strict timing budgets and focus solely on inference rather than policy reasoning. Their impact is reflected in the KPMs observed at the Near-RT layer, as illustrated in Fig.~\ref{fig:flow}. Shifts in channel conditions or interference can degrade the effectiveness of a deployed WPFM, triggering feedback to the upper layers and, if needed, model adaptation.


The interaction cycle becomes fully agentic once model governance is added. When the LLM rApp detects degrading KPMs that cannot be explained by configuration conflicts or by the current A1 policy, it can infer that an underlying model has become stale or misaligned with current conditions. It then triggers retraining or fine-tuning via the MCP-aware MLOps pipeline, which retrieves historical data, trains candidate models, and publishes updated versions to the Model Catalog. Crucially, the LLM rApp is not restricted to a single replacement. It can select the best-performing candidate or temporarily fall back to a simpler AI/ML or legacy rule-based xApp to maintain acceptable performance. After deployment, subsequent telemetry is used to confirm recovery; otherwise, the rApp can adjust policies, switch to an alternative model, or roll back to a previously validated configuration. 

Through this collaboration of agents, data streams, and learning mechanisms, the system supports a continuous cycle of perception, action, and correction across timescales. Standard O-RAN interfaces and service models bind these agents together and ensure that model governance remains grounded in measured network behavior and cross-loop decisions remain coordinated.

\subsection{Safety, Guardrails, and Conflict Resolution}
Although agentic AI enables unprecedented autonomy and adaptability in O-RAN, it also introduces new operational risks that must be addressed explicitly. Unlike conventional xApps that implement narrowly scoped control logic, agentic systems can generate policies, reconfigure control applications, and alter the deployed model set itself. Without appropriate safeguards, this autonomy can create unstable feedback loops, unintended cross-layer interactions, or violations of operational and regulatory constraints. Therefore, safety and governance shall be treated as equally crucial as architectural components.

At the strategic layer, safety begins by constraining the reasoning and action space of the LLM rApp. The agent is never permitted to issue unconstrained actions. All A1 policies are generated using structured schemas and are validated against hard constraints stored in the Knowledge Base, including spectrum regulations, transmission power limits, slice isolation rules, and operator-defined service-level objectives. In addition, before any new policy or model is deployed to the Near-RT RIC, the LLM rApp uses the Simulation/DT block in the SMO to evaluate expected impact using historical data. This validation helps detect unsafe behaviors and policy side effects early, without exposing live users to instability.

Strategic-layer safety also applies to model governance. The LLM rApp monitors long-horizon KPM trends via O1 telemetry and the Analytics module, but it does not treat every regression as a retraining event. Instead, it considers alternative explanations, such as intent misconfiguration or conflicting xApp behavior, and prioritizes the least disruptive corrective action. In many cases, this may involve adjusting A1 policy parameters or switching between existing xApps rather than deploying a newly trained model. Only when these measures fail does the agent trigger fine-tuning or retraining through the MLOps pipeline, and any candidate model must pass offline validation before deployment.

Within the Near-RT RIC, safety is primarily a coordination problem. Multiple xApps may attempt to influence overlapping parameters, and their interactions can create oscillations or inconsistent control. In the proposed architecture, SLM xApps act not only as controllers but also as orchestrators of other AI/ML and legacy xApps. 
To keep this flexibility bounded, the Near-RT RIC enforces explicit decision constraints, including rate limits, parameter bounds, and stability guards that prevent rapid oscillations and unsafe updates. When multiple xApps propose incompatible actions, arbitration logic resolves conflict based on policy priority, confidence estimates, and observed performance impact, yielding deterministic behavior even under concurrent control.

At the RT layer, safety is governed by two hard requirements: strict timing constraints and bounded actuation. WPFM-based dApps execute within the O-DU pipeline and must meet execution deadlines. Any inference that exceeds its time budget is discarded, and control reverts to deterministic baseline logic, such as a legacy scheduler or vendor-certified PHY procedures. Beyond timing checks,  WPFM outputs are subject to sanity constraints to ensure that the suggested link-adaptation and beam-management actions remain within standard-compliant bounds.


Safety across all layers is reinforced through explicit rollback mechanisms. Policies, xApp configuration, and model version are tracked through the Model Catalog and MLOps pipeline. If telemetry indicates sustained KPM degradation following a change, the system can rapidly revert to a previously validated safe configuration or a deterministic baseline to restore stability, even if that configuration is not optimal under the new environment. 
Together, bounded tactical control, deterministic real-time fallbacks, and continuous rollback provide a practical safety envelope for agentic operation, allowing the system to adapt while remaining operationally predictable.



\section{Proof of Concept}
\label{sec:implementation} 
To demonstrate the feasibility of the proposed agentic AI framework within current O-RAN constraints, we implement a proof-of-concept system spanning the Non-RT and Near-RT layers and emulating key aspects of the RT intelligence loop. The implementation combines a live 5G RAN based on srsRAN, an O-RAN-SC Near-RT RIC, and a set of lightweight AI agents that realize the strategic and tactical roles described in Section~\ref{sec:framework}. Because a fully standardized and openly deployable RT RIC is not yet available, WPFM behavior is evaluated off-path using recorded IQ data. This approach allows us to study agentic model governance, performance degradation, and retraining decisions without inserting inference directly into the real-time DU processing chain. 

We present this proof of concept through two complementary use cases that highlight different aspects of agentic intelligence. The first focuses on agentic model governance, whereas the performance of a WPFM degrades under changing radio conditions, prompting the LLM governor to decide whether to retrain or fine-tune the model. The second use case demonstrates intent-driven slice resource control in a live network with four slices. Here, an SLM-based xApp performs slice-level allocation in the Near-RT loop, while the LLM supervises policy objectives, updates constraints, and adjusts priorities based on observed KPMs. 


\subsection{Use Case I: Agentic WPFM Governance and Retraining}

The foundation model is pre-trained in a self-supervised manner using the FFT of I/Q samples. In this work, we use our previously published wireless foundation model \cite{cheraghiniaFoundationModelWireless2025} and adapt it to new downstream scenarios that can be deployed at the DU level. We assume that LLM automates the overall workflow and, in some cases, identifies the need to fine-tune the WPFM in response to recent network announcements or changes, far surpassing the limitations of manual operator-driven monitoring and model adjustments. To demonstrate this process, we consider two evaluation scenarios.

In the first scenario, the downstream task is wireless technology recognition, which can help with load balancing in vehicular applications (including LTE, 5G-NR, Wi-Fi, ITS-G5, and C-V2X)\footnote{https://ieee-dataport.org/documents/dataset-iq-samples-lte-5g-nr-wi-fi-its-g5-and-c-v2x-pc5}. The LLM processes new announcements and initiates fine-tuning of the WPFM to extend the number of supported classes and environments using newly available data stored in the data lake. In the second scenario, we assume that interference between LTE and 5G NR leads to degradation in KPM\footnote{https://ieee-dataport.org/documents/dataset-fft-iq-samples-lte-nr-and-overlap-0}. In response, the LLM, together with an SLM, detects the KPM degradation and triggers a new downstream task for interference detection using the WPFM.

\begin{figure}
    \centering
    \includegraphics[width=\linewidth]{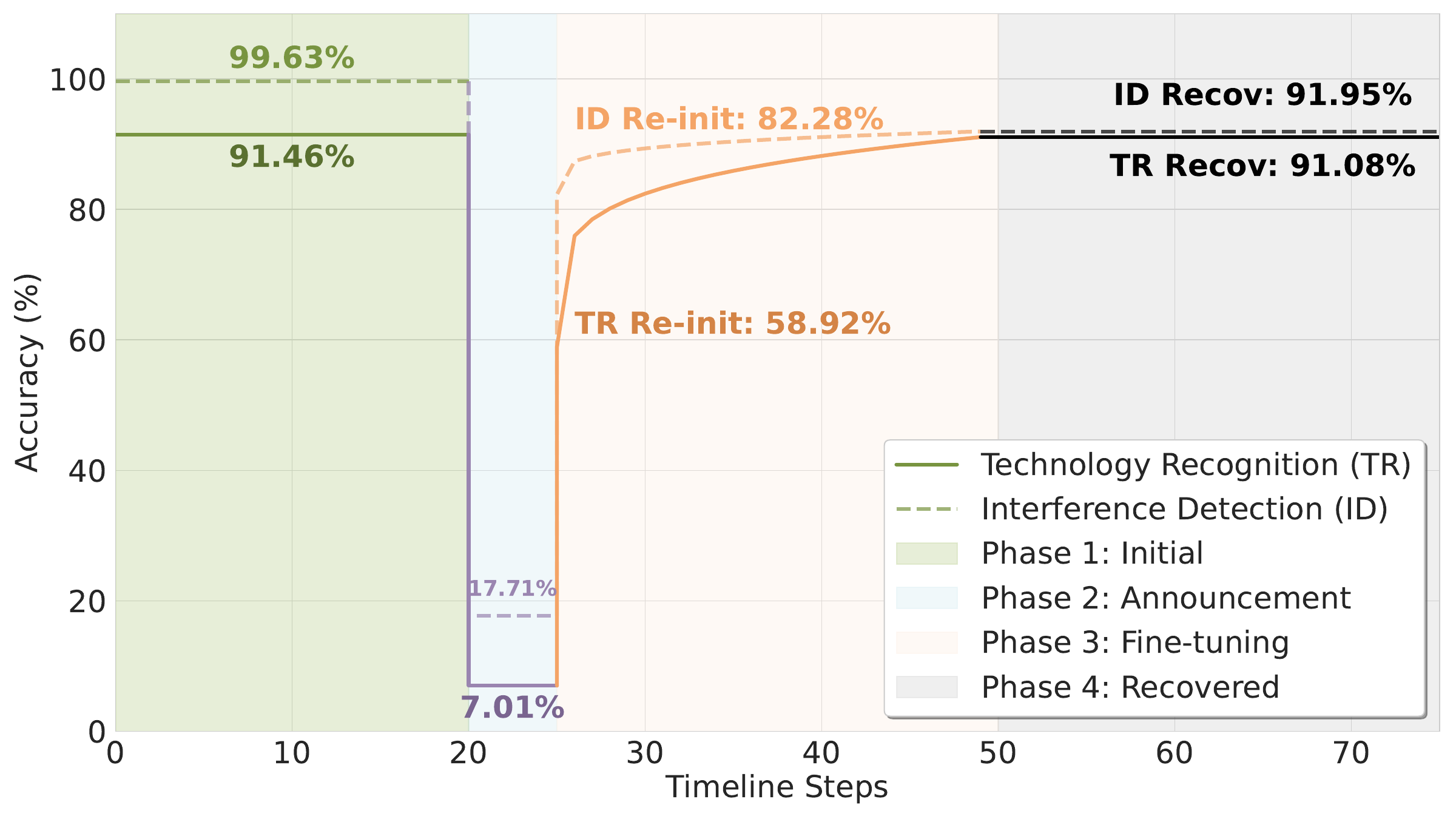}
    \caption{WPFM governance for two scenarios: a) New class announcement for Technology Recognition (TR) in vehicular application. b) Interference Detection (ID) announcement for 5G-NR and LTE}
    \label{fig:WPFM-tuning}
\end{figure}

To illustrate this process, we consider four stages of the WPFM fine-tuning procedure, as shown in Fig.~\ref{fig:WPFM-tuning}. In the 'initial' stage, the model operates with high accuracy before any new announcements or changes are introduced, and the same model can be used until the new model is trustworthy enough to be deployed. During the 'announcement' stage, new requirements for downstream tasks are introduced, causing a temporary drop in model performance. In the 'fine-tuning' stage, the model begins the training process using data from the data lake. After the first fine-tuning epoch, corresponding to one time step in the figure and taking approximately 17 seconds, the model achieves accuracies of 58.92\% and 82.28\%. After approximately 340 seconds, the model is fully fine-tuned and can be redeployed into the network, supporting the new classes.

\subsection{Use Case II: Agentic Slice Resource Allocation}
This use case is validated on a live 5G setup based on srsRAN, with four active slices: two bandwidth-intensive slices, one latency-sensitive slice, and a fourth slice initially configured as a regular service that is later updated to a VIP slice via an operator intent prompt. The VIP slice is a throughput-oriented premium service. Downlink traffic is generated using heterogeneous traffic patterns and time-varying load profiles to induce congestion and non-stationary behavior. The traffic scenario includes stepwise additions and removals of UDP flows with different inter-departure and packet-size distributions across slices, creating realistic fluctuations in offered load and queue dynamics. 

In the Non-RT RIC, an LLM agent (Nvidia Nemotron on an H200 GPU) translates operator intent into a structured policy that defines control guardrails. In the Near-RT RIC, an SLM agent (GPT-OSS on an RTX 5090 GPU) makes low-latency Physical Resource Block (PRB) allocation decisions using live KPMs that respect policy constraints. System monitoring, control, and human-in-the-loop interaction are handled through CLI-LLM \cite{cli-llm-paper}, which connects live telemetry to the agents' prompts and exposes control endpoints for runtime intent updates.

In this use case, the Near-RT SLM xApp implements slice-level PRB allocation using a 5-second sliding window of E2SM-KPM measurements updated every second. The input KPMs include downlink throughput, resource efficiency, buffer occupancy, and RLC delay, which are jointly used to compute per-slice PRB caps under congestion. 

The Non-RT LLM rApp supervises this controller at a coarser timescale by analyzing one-minute KPM summaries to detect persistent starvation, oscillatory behavior, or persistent latency violations. Rather than issuing direct control actions, this agent periodically updates a structured policy object that specifies service intent, minimum resource floors, and maximum per-step adjustment limits. This policy constrains the SLM agent's fast decision loop and preserves stability.

Runtime intent changes are injected through operator prompts. During the scenario, slice 4 is promoted to a VIP service while the network is under load. The LLM agent incorporates this updated intent into the policy object, increasing the priority of slice 4 while maintaining explicit protection for the latency-sensitive slice. The SLM then enforces these updated constraints in real time without requiring retraining or controller restarts. 

\begin{figure*}[b]
    \centering
    \includegraphics[width=\linewidth]{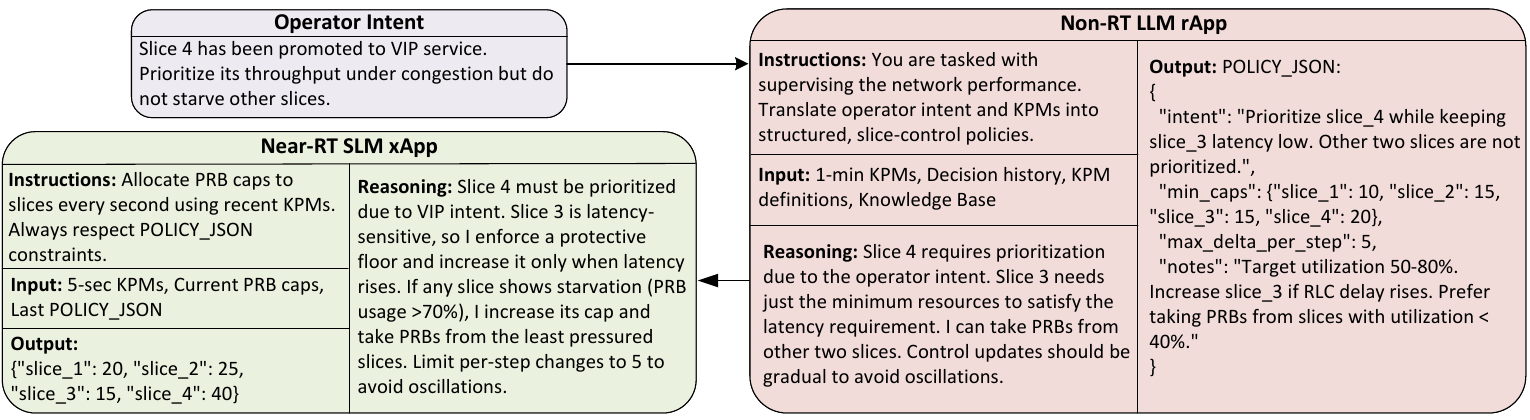}
    \caption{Representative examples of agentic interaction illustrating how telemetry inputs and operator intent are processed through distinct reasoning stages to generate structured policies and control actions.}
    \label{fig:chat}
\end{figure*}

Figure \ref{fig:chat} provides illustrative examples of policy structures and message exchanges for this interaction. The reasoning snippets are abbreviated for readability and are not full execution logs. After the VIP promotion, the LLM updates the policy, and the SLM immediately reflects this by shifting available PRBs toward slice 4 while keeping the low-latency slice at its protective floor.

\begin{figure*}
    \centering
    \includegraphics[width=\linewidth]{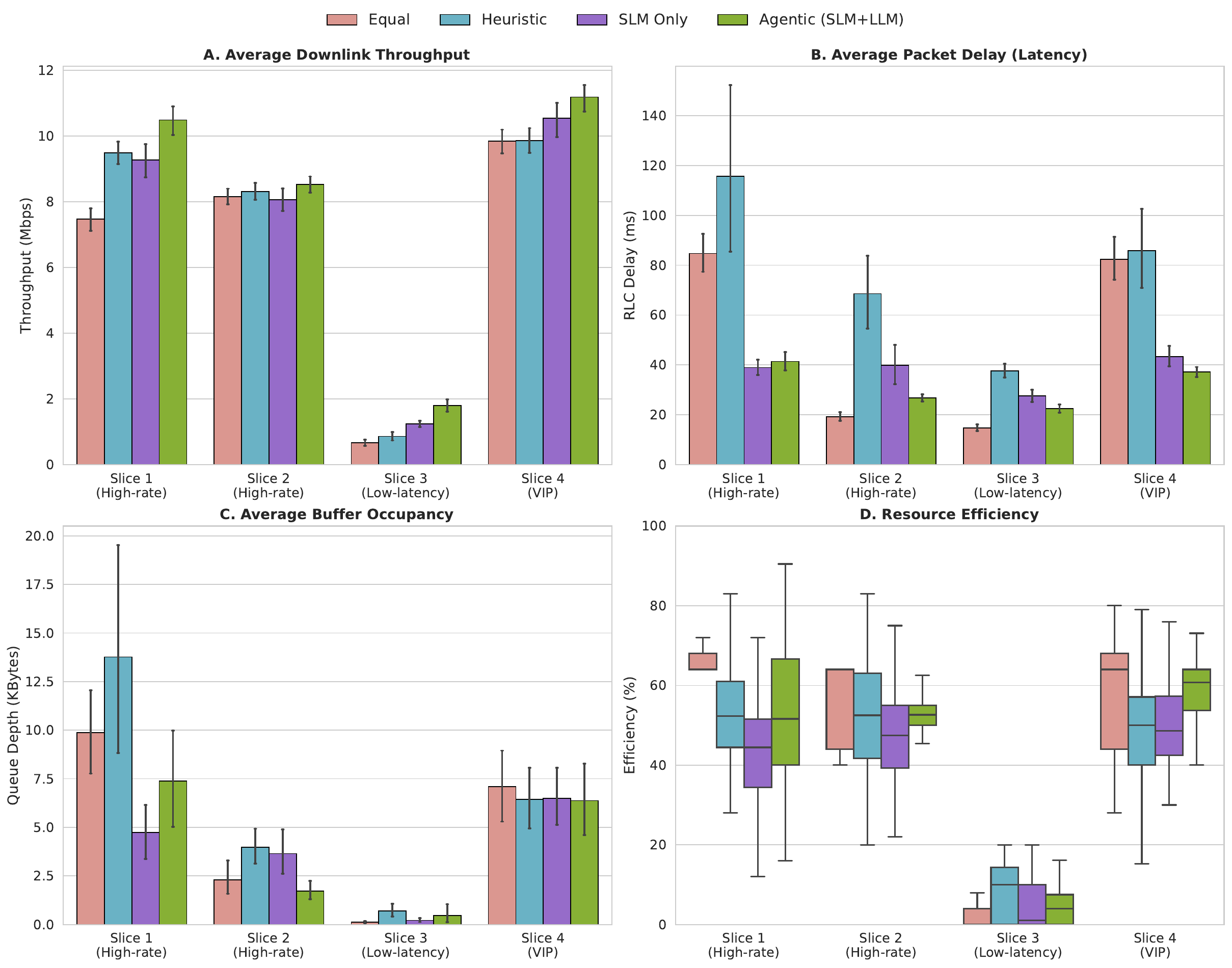}
    \caption{Comparison of static, heuristic, SLM-only, and agentic slice control strategies in a live 5G setup.}
    \label{fig:rrm_results}
\end{figure*}

To evaluate the benefits of the agentic approach, we compare four control strategies: static equal PRB allocation, a heuristic slice controller, an SLM-only controller with fixed objectives, and the proposed agentic controller combining SLM execution with LLM supervision. The heuristic baseline allocates resources based on short-term demand and utilization feedback, computing a per-slice priority score from KPM-derived indicators. Performance is assessed based on downlink throughput, radio latency for the latency-sensitive slice, buffer occupancy, and overall resource efficiency, with particular attention to the system's response following the VIP promotion of the fourth slice.

As depicted in Fig.~\ref{fig:rrm_results}, the static and heuristic schemes struggle to adapt to the combined effects of congestion and changing service priorities. While the heuristic controller improves average throughput over static allocation, it exhibits a delayed response to intent changes and greater delay variance.

The SLM-only controller reacts more quickly to short-term KPM fluctuations, increasing VIP throughput to 10.5 Mbps, but it lacks awareness of higher-level service priorities. As a result, it trades VIP gains for latency protections, leading to higher delays for the latency-sensitive slice than under agentic control. 

In contrast, the proposed agentic controller achieves stronger alignment with service intent while maintaining stability. Following the promotion of slice 4 to VIP status, the LLM-supervised system rapidly adjusts policy floors and priorities, enabling the SLM to increase the average VIP throughput by approximately 6\% relative to the SLM-only baseline while preventing starvation of the other slices. At the same time, the policy-enforced minimum caps and tighter step-size limits preserve an average low latency of 22 ms for the latency-sensitive slice. Across all phases, the agentic approach achieves higher VIP throughput, comparable or lower delay, and improved overall resource efficiency, defined as the ratio of PRBs actually used to the total number of PRBs allocated.




\section{Risks \& Challenges}
\label{sec:challenges} 
While the proposed agentic AI framework offers a compelling path toward autonomous O-RAN operation, several technical, architectural, and operational challenges must be addressed before such systems can be deployed at scale.

The first challenge lies in the real-time feasibility of learning-based control, particularly in the lower layers of the RAN. Although WPFM-based dApps promise rapid inference from IQ or CSI data, meeting sub-millisecond timing constraints in practical O-DU implementations remains non-trivial. Hardware acceleration, memory locality, and deterministic execution pipelines are essential, yet not uniformly available across vendor platforms. Moreover, the absence of a fully standardized RT RIC within O-RAN means that assumptions about dApp placement and execution remain provisional. While early proposals advocate embedding RT intelligence directly within the O-DU processing chain, broader ecosystem alignment and reference implementations are still evolving \cite{lacavaDAppsEnablingRealtime2025}.

A second challenge concerns data availability and semantic alignment across layers. Agentic reasoning relies on consistent, semantically meaningful telemetry flowing among the RT, Near-RT, and Non-RT domains. In practice, O-RAN exposes heterogeneous data streams with different sampling rates, delays, and levels of abstraction. Mapping PHY‑level observations to high‑level intent, or correlating long‑horizon KPMs with instantaneous radio events, remains an open problem. Furthermore, public datasets rarely provide synchronized KPM and IQ data, limiting the ability to validate cross-layer agentic behavior without simulations or a sandbox environment. 

The robustness of learning-based agents under non-stationary conditions presents another significant risk. Wireless environments are inherently dynamic, influenced by mobility, interference, weather, and user behavior. As a result, SLM and WPFM models are subject to concept drift and performance degradation over time. While the proposed framework includes mechanisms for drift detection, retraining, and rollback, determining when to retrain, which model to deploy, and how to avoid oscillations between models remain open research challenges. Overly aggressive retraining can destabilize control loops, while conservative policies may delay adaptation to genuine environmental changes \cite{poleseUnderstandingORANArchitecture2023a}.

Trust and explainability represent additional hurdles, particularly for operator-facing systems. Although LLMs enable high-level reasoning and intent translation, their internal decision processes are not inherently transparent. Operators may be reluctant to cede control to systems whose recommendations cannot be easily audited or justified. This challenge is intensified when agentic systems orchestrate other xApps or override legacy control logic. Providing interpretable policy rationales, confidence estimates, and traceable decision histories will be essential to building operational trust and to satisfying regulatory and operational accountability requirements \cite{brikExplainableAI6G2025}. Promising research directions include neuro-symbolic approaches that combine data-driven learning with explicit constraints and rules. 

Finally, standardization and interoperability remain open issues. While A1, E2, and O1 interfaces provide a solid foundation, current O-RAN specifications were not designed with agentic AI or dynamic model governance in mind. Expressing complex intents, representing model confidence, or negotiating control authority among multiple agents may require extending existing service models or introducing new abstractions. Without careful alignment between research prototypes and evolving standards, there is a risk that agentic frameworks remain confined to experimental deployments rather than transitioning into interoperable, multi-vendor systems \cite{poleseUnderstandingORANArchitecture2023a}.

Despite these challenges, none of them represent fundamental barriers. Instead, they emphasize the importance of taking a layered, cautious, and standards-aware approach to agentic intelligence in O-RAN. By explicitly acknowledging these risks and incorporating safeguards, fallback mechanisms, and human oversight into the architecture, agentic AI can be introduced to augment existing control mechanisms while preserving the expected reliability and predictability.

\section{Conclusion}
\label{sec:conclusion} 

This paper has proposed a multi-scale agentic AI framework for O-RAN that closes the gap between the rising complexity of 6G RAN and the narrow, siloed intelligence in current deployments. Rather than treating O-RAN control applications as independent optimization modules, our proposed architecture defines a coordinate hierarchy of autonomous actors across different timescales. Through a proof-of-concept implementation, we have demonstrated how intent supervision, tactical control, and real-time inference can be coordinated in practice.  
Although challenges remain, particularly RT standardization gaps, cross-layer data alignment, and verifiable agent behavior, the framework outlines a practical path to introduce agentic intelligence incrementally and safely toward autonomous O-RAN operation in 6G networks.

\section*{Acknowledgement}
This work was supported by the Horizon Europe program under the MCSA Staff Exchanges 2021 grant agreement 101086218 (EVOLVE project) and the Horizon-JU-SNS-2023 Research and Innovation Program under Grant Agreement No. 101139194 (6G-XCEL project).



\bibliographystyle{myIEEEtran}
\bibliography{ref.bib}

\end{document}